\documentclass{aa}
\usepackage{graphicx}
\usepackage{amssymb}
\usepackage{latexsym}

\begin{document}

\title{A NICMOS search \\for obscured Supernovae in starburst galaxies
	\thanks{Based on observations made with the NASA/ESA Hubble Space Telescope associated with program 9726, 
	obtained at the Space Telescope Science Institute, which is operated by the Association of Universities 
	for Research in Astronomy, Inc., under NASA contract NAS 5-26555,  
	and on data obtained at the VLT through the ESO program 272.D-5043.
	}}

\author{G. Cresci \inst{1,2} \and F. Mannucci \inst{3} \and M. Della Valle \inst{2,4}  \and R. Maiolino \inst{1}}

\institute{Max-Planck-Institut f\"ur extraterrestrische Physik,
        Postfach 1312, D-85741 Garching, Germany
		\and INAF, Osservatorio Astrofisico di Arcetri, 
	Largo E. Fermi 5, I-50125, Firenze, Italy  
		\and Istituto di Radioastronomia, INAF, 
	Largo E. Fermi 5, I-50125, Firenze, Italy
		\and Kavli Institute for Theoretical Physics
		University of California
		Santa Barbara, CA   93106 
	}

\offprints{G. Cresci \\ 
	\email{gcresci@mpe.mpg.de}}

\date{Received / Accepted}

\abstract{ The detection of obscured supernovae (SNe) in near-infrared
    monitoring campaigns of starburst galaxies has shown that a
    significant fraction of SNe is missed by optical surveys. However,
    the number of SNe detected in ground-based near-IR observations is
    still significantly lower than the number of SNe extrapolated from
    the FIR luminosity of the hosts. A possibility is that most SNe
    occur within the nuclear regions, where the limited angular
    resolution of ground-based observations prevents their
    detection. This issue prompted us to exploit the superior angular
    resolution of NICMOS-HST to search for obscured SNe within the
    first kpc from the nucleus of strong starbursting galaxies. A
    total of 17 galaxies were observed in SNAPSHOT mode. Based on
    their FIR luminosity, we expected to detect not less than $\sim
    12$ SNe. However, no confirmed SN event was found. 
    From our data we derive an observed nuclear SN rate $ <0.5$ SN/yr
    per galaxy. The shortage of SN detections can be explained by a
    combination of several effects. The most important are: i) the
    existence of a strong extinction, $A_V\gtrsim 11$; ii) most SNe
    occur within the first $0.5\arcsec$ (which corresponds in our
    sample to about 500pc) where even NICMOS is unable to detect SN
    events.  
}

\maketitle


\section{Introduction}

Current measurements of supernova (SN) rate are entirely based on
events detected during surveys carried out at the optical
wavelengths. This constitutes a problem, since many SNe could be 
obscured by dust. As a consequence, the derived SN rates, especially for 
core--collapse and Ia ``prompt'' SNe (Mannucci et al. \cite{mannucci05}, 
\cite{mannucci06}) in the distant universe, may be only lower limits.  
As an example, we point
out that several monitoring campaigns of starburst galaxies, carried
out at the optical wavelengths, do not show evidences for an enhanced
SN rate with respect to normal quiescent galaxies (e.g. Richmond et
al. \cite{richmond}, Navasardyan et al. \cite{navasardyan}). 
Near-infrared (near-IR) observations can shed light on this problem, 
as the extinction at these wavelengths is about ten times lower
than in the optical. The importance of near-IR was demonstrated by Maiolino et
al. (\cite{maiolino}), who showed, on the basis of a small sample of SNe
detected in the near-IR, that the rate of supernovae measured in the
optical could be significantly reduced by dust extinction up to an order of 
magnitude. 

During the last decade other near-IR searches for extincted SNe were
attempted.  Van Buren et al.  (\cite{vanburen}) detected
SN1992bu (a possible SN not confirmed spectroscopically).  Grossan et
al. (\cite{grossan}) monitored a large number of galaxies over two years, but
failed to detect extinguished SNe, probably because of their
poor spatial resolution. Bregman et al. (\cite{bregman}) searched for line emission  
at longer wavelengths (6.63 $\mu m$ using ISOCAM), but did not detected 
any feature connected with SN events. More recently, 
Mattila et al. (\cite{mattilaSN2}, \cite{mattilaSN3}) discovered 
two SNe during their monitoring in Ks band with the William Herschel Telescope. 
Di Paola et al. (\cite{dipaola}) reported the near-IR
serendipitous discovery of SN2002cv, an high extincted ($A_V\sim 8$)
type Ia SN. \\ 
In late 1999 Mannucci et al. (\cite{mannucci}) started a K'-band 
monitoring campaign
of a sample of 46 Luminous Infrared Galaxies (LIRGs, $L_{FIR} >
10^{11}\ L_{\sun}$), aimed at detecting obscured SNe in the most 
powerful starbursting galaxies. 
During the monitoring 4 SNe were detected, two of which were
discovered by our group: SN1999gw (Cresci et al. \cite{cresci}) and
SN2001db, the first SN detected in the near-IR which has received a
spectroscopic confirmation (Maiolino et al.  \cite{maiolino}).
\begin{table*}	
\label{sample}
	\begin{center}
	\begin{tabular}{l c c c c}
	\hline \hline
	Galaxy & \multicolumn{2}{c}{R.A. \ \ (J2000)\ \  DEC.} & $\mathcal{L}_{FIR}$ & z  \\
	\hline
	NGC 34	        & 00h11m06.5s & $-$12d06m26s & 11.43 & 0.020 \\
	NGC 1614        & 04h33m59.8s & $-$08d34m44s & 11.41 & 0.016 \\
	VII-ZW031       & 05h16m47.3s & $+$79d40m12s & 11.94 & 0.053 \\
	IRAS 05189-2524 & 05h21m01.4s & $-$25d21m45s & 11.89 & 0.042 \\
	IRAS 08572+3915 & 09h00m25.4s & $+$39d03m54s & 11.99 & 0.058 \\
	UGC 5101        & 09h35m51.4s & $+$61d21m11s & 11.90 & 0.039 \\
	NGC 3256        & 10h25m51.8s & $-$43d54m09s & 11.44 & 0.009 \\
	IRAS 10565+2448 & 10h59m18.1s & $+$24d32m34s & 11.87 & 0.042 \\
	NGC 3690        & 11h28m31.9s & $+$58d33m45s & 11.72 & 0.011 \\
	NGC4418         & 12h26m54.6s & $-$00d52m39s & 11.00 & 0.007 \\
	Mrk 273         & 13h44m42.1s & $+$55d53m13s & 12.10 & 0.038 \\
	UGC 8782        & 13h52m17.7s & $+$31d26m44s & 12.27 & 0.045 \\
	IRAS 14348-1447 & 14h37m38.2s & $-$15d00m26s & 12.27 & 0.082 \\
	Arp 220         & 15h34m57.3s & $+$23d30m12s & 12.12 & 0.018 \\
	NGC 6090        & 16h11m40.3s & $+$52d27m26s & 11.35 & 0.029 \\
	IRAS 20414-1651 & 20h13m29.4s & $-$16d40m16s & 11.99 & 0.087 \\
	IRAS 23128-5919 & 23h15m46.8s & $-$59d03m14s & 11.80 & 0.044 \\
	\hline \hline
	\end{tabular}
	\caption{The galaxy sample.  The RA and DEC position reported 
	correspond to the optical center of the galaxies; 
	$\mathcal{L}_{FIR}$ is $\log(L_{FIR}/L_{\sun})$ (Mannucci et al. 
	\cite{mannucci}, where $L_{FIR}$ is defined accordingly to Helou 
	et al. \cite{helou88} as the luminosity between 42.5 to 122.5
	$\mu m$); \textit{z} is the redshift.} 
	\end{center}
\end{table*}

The number of detected events was about an order of magnitude higher
than expected from the B band luminosity of the parent galaxies
(Cappellaro et al. 1999), showing that the B band luminosity can not be used to trace 
the star formation in starburst galaxies.  Although these results highlight the
capability of near-IR observations to reveal obscured SNe, otherwise
missed in classical optical surveys, Mannucci et al. (2003) also showed
that the inferred SN rate was still 3-10 times lower than that expected from the
far-infrared (FIR) luminosity of the galaxies of the sample.
A possible explanation for the shortage of near-IR SNe is the presence
of dust extinction $\rm A_V > 25$~mag, which would make SNe heavily obscured even in
the near-IR. Another reason could be the reduced capability of
ground-based observations to detect SNe within the first $2\arcsec$
from the galactic nuclei. The reduced sensitivity in the nuclear
region is a direct consequence of the presence of the residuals, which
are obtained during the images subtraction (see Mannucci et
al. \cite{mannucci} for details). Even if $2''$ corresponds to a
relatively small region for the host, it could contain most of the
star-formation activity and produce most of the near-IR flux (Soifer et
al. \cite{soifer00}, Soifer et al. \cite{soifer01}). If most SNe are
really lost in the nuclear regions, one can explain the shortage of
SN detections without calling for a large near-IR extinction.\\
In this paper we present a near-IR search for nuclear SNe 
exploiting the superior angular resolution of NICMOS-HST. 
The capabilities of NICMOS to search for nuclear obscured SNe, 
the observations and data reduction will be presented in the following section; 
in Sect.~\ref{expectedsnsect} we derive the expected number of SNe in our data, 
that is compared with the results obtained in Sect.~\ref{implications}; 
our conclusions follow in Sect.~\ref{conclusionsection}.
\begin{table*}	
	\begin{center}
	\begin{tabular}{ @{} l c c c c c c c c c @{}} 
	\hline \hline
	Galaxy & \multicolumn{2}{c}{First Epoch} & & \multicolumn{2}{c}{Second Epoch} & & \multicolumn{3}{c}{Comparison} \\ 
		\cline{2-3} \cline{5-6} \cline{8-10}
	       & Date & Exp. Time & & Date & Exp. Time & & $\textrm{Mag}_L(in)$ & $\textrm{Mag}_L(out)$ & Non-Det. \\
	       &      &    (s)    & &      &    (s)    & &                      &                       & Radius \\
	\hline
	NGC 34   	& 1997 Dec 2  & 255.98 & & 1998 Oct 11 & 255.98 & & 17.8 & 20.2 & $0.45\arcsec$\\
	NGC 1614        & 1998 Feb 07 & 383.85 & & 2004 Gen 05 & 599.44 & & 17.0 & 20.7 & $0.68\arcsec$\\
	VII-ZW031       & 1997 Nov 17 & 599.44 & & 2003 Dec 31 & 599.44 & & 19.0 & 21.5 & $0.52\arcsec$\\
	IRAS 05189-2524 & 1997 Dec 09 & 223.76 & & 2003 Sep 29 & 599.44 & & 17.6 & 21.4 & $0.38\arcsec$\\
	IRAS 08572+3915 & 1997 Nov 11 & 159.81 & & 2004 Mar 17 & 599.44 & & 20.2 & 20.7 & $0.30\arcsec$\\
	UGC 5101        & 1997 Nov 07 & 559.48 & & 2004 Feb 07 & 599.44 & & 18.6 & 21.3 & $0.52\arcsec$\\
	NGC 3256        & 1997 Nov 28 & 191.83 & & 2003 Nov 17 & 599.44 & & 18.3 & 20.8 & $0.30\arcsec$\\
	IRAS 10565+2448 & 1997 Nov 29 & 479.54 & & 2003 Nov 11 & 599.44 & & 19.1 & 21.2 & $0.45\arcsec$\\
	NGC 3690 - N1   & 1997 Nov 04 & 207.80 & & 2003 Sep 02 & 599.44 & & 19.0 & 20.8 & $0.45\arcsec$\\
	NGC 3690 - N2   & 1997 Nov 29 & 207.80 & & 2003 Sep 02 & 599.44 & & 18.5 & 20.7 & $0.52\arcsec$\\
	NGC 4418        & 1997 Nov 26 & 351.66 & & 2004 Apr 26 & 599.44 & & 19.1 & 21.0 & $0.60\arcsec$\\
	Mrk 273         & 1997 Dec 10 & 255.73 & & 2004 May 11 & 599.44 & & 18.8 & 21.0 & $0.38\arcsec$\\
	UGC 8782        & 1998 Aug 19 & 2175.8 & & 2004 Mar 17 & 599.44 & & 19.6 & 21.9 & $0.52\arcsec$\\
	IRAS 14348-1447 & 1998 Jul 09 & 479.54 & & 2004 Apr 23 & 599.44 & & 19.5 & 21.2 & $0.45\arcsec$\\
	Arp 220         & 1997 Apr 04 & 1023.0 & & 2004 Jan 10 & 599.44 & & 19.0 & 21.2 & $0.35\arcsec$\\
	NGC 6090        & 1997 Nov 10 & 383.63 & & 2003 Dec 01 & 599.44 & & 18.8 & 21.3 & $0.45\arcsec$\\
	IRAS 20414-1651 & 1998 Jul 01 & 639.85 & & 2003 Oct 24 & 599.44 & & 21.3 & 21.5 & $0.45\arcsec$\\
	IRAS 23128-5919 & 1998 Mar 08 & 639.85 & & 2003 Oct 13 & 599.44 & & 18.9 & 21.8 & $0.52\arcsec$\\
	\hline
	\hline
	\end{tabular}
	\caption{\label{observ} 
	Summary of the first epoch archive image and second epoch new NICMOS observation.  
	\textit{Date} is the date on which this image was taken, \textit{Exp. Time} is the 
	exposure time in second used for that image. 
	\textit{$\textrm{Mag}_L$} the limiting magnitude obtained in the subtraction in the region of the 
	galaxy interested by the subtraction residuals (\textit{in}) and in the outer parts 
	(\textit{out}); \textit{Non-Det. Radius} the radius of the central region where the detection 
	of SNe is not possible.}
	\end{center}
\end{table*}

\section{The use of NICMOS to search for nuclear obscured SNe}

The HST near-IR camera NICMOS is an excellent tool to search for obscured nuclear SNe in starburst 
galaxies, since it joins a high sensitivity in the near-IR, where the dust extinction is greatly 
reduced, to a high angular resolution ($0.2\arcsec$) and stable PSF, allowing for a much easier 
detection of SNe close to the nucleus in the difference image. \\

We selected a sample of 35 Luminous and Ultraluminous Infrared
Galaxies (LIRGs and ULIRGs) closer than about 450 Mpc ($z \leq 0.1$)
and already observed once with NICMOS in the F160W filter with the
NIC2 camera, so that a first epoch image was already available.  We
then obtained a second epoch in SNAPSHOT mode for 16 of the selected
galaxies. For one galaxy, NGC 34, we have compared two images at
different epochs already present in the archive (see
Tab.~\ref{observ}) The list of these 17 galaxies is given in
Tab.~\ref{sample}, along with their far-IR flux and redshift, while the
relative observational setups are given in Tab.~\ref{observ}.  Images were
reduced by using the \textit{On The Fly Reprocessing} system of the
HST archive (Swam et al. \cite{swam}).

\begin{figure}
	\centering
	\includegraphics[width=0.45\textwidth]{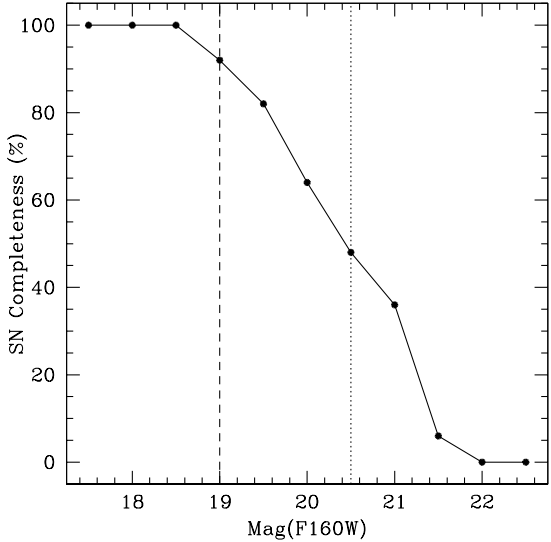}
	\caption{\label{arp220simul} Example of SN recovering simulations  
	for the circular regions 1.35$\arcsec$ around the nucleus in ARP 220.
	Although some SN are lost in correspondence with the strongest residuals 
	at the most conservative $\textrm{Mag}_L(in)=19.0$ (dashed line), corresponding to the 90\% 
	recovering completeness in this region, much dimmer sources are still detectable at 
	different locations, such as the possible SN discussed in Appendix which was detected  
	with $\mathrm{H}=20.5$ (dotted line).}
\end{figure}

The comparison of the images was in fact performed with ISIS, a tool developed by 
Alard \& Lupton (\cite{alard1}) and refined by Alard (\cite{alard2}).  
The images are aligned using field stars, than the image with the best PSF is 
selected as reference to be compared with all others frames of the galaxy. In order to 
correct the effects of the variable PSF, the reference image is convolved with an appropriate 
kernel determined by a least square fit. The images are then normalized in flux and subtracted. 
Typically we were able to subtract $98\%$ of the galaxy flux.\\
Although the NICMOS PSF is much more stable than in
ground based images, the results of the subtraction of NICMOS images
are still affected by residuals in the nuclear regions, but on a smaller scale. 
The residuals are brighter where the emission has a 
strong radial gradient and non circular structures (such as the diffraction spikes of the spider 
arms that are not well reproduced even by the complex ISIS kernel). 
The presence of residuals is due to the HST PSF variations over long time
scale (e.g. Krist \& Hook \cite{krist97}) and to the different
orientation of the diffraction spikes due to the different roll angle
of HST at the various epochs of observation. As a consequence, the
limiting magnitude for point sources detection is much 
brighter than expected from the photon noise and it is strongly
dependent on the location and on the distance from the galactic center, as the subtraction
residuals are concentrated in the nuclear regions of the galaxies.\\
For each galaxy we have evaluated two different limiting magnitudes,
$\textrm{Mag}_L(in)$, in the region of the galaxy affected by the
residuals of the subtraction (between $0.5\arcsec$ and $1.35\arcsec$
from the nucleus, on average) and $\textrm{Mag}_L(out)$ in the outer
parts. The SN detection limit was estimated through simulations, by
adding artificial stars to the original images before image subtraction.  
The artificial stars were added at random locations,  
respectively inside and outside the circular aperture around the nucleus 
were significant residuals of the subtraction 
are visible, and with different luminosities. The inferred limiting
magnitudes for SN detection, defined as the completeness level of
$90\%$, are listed in Tab.~\ref{observ}. 
However, the limiting magnitude is strongly dependent from the particular location  
inside the circular area around the nucleus, due to the presence of stronger 
residuals corresponding to the diffraction spikes or to secondary nuclei 
of the galaxies. As an example, in Fig.~\ref{arp220simul} we report the results 
of our SN recovering completeness simulations inside the circular area around the nucleus 
in ARP 220. Although some SNe are lost in correspondence with the strongest residuals 
at the most conservative $\textrm{Mag}_L(in)=19.0$, much dimmer sources are still detectable at 
different locations, such as the possible SN observed in this galaxy 
which was detected with $\mathrm{H}=20.5$ (see Appendix).\\
In the following the number of
expected SNe will be evaluated using both $\textrm{Mag}_L(out)$ and
the more conservative limiting magnitude $\textrm{Mag}_L(in)$,
relative to the nuclear regions of the galaxies where most of the
starburst activity are expected to be hosted. \\
Furthermore the ISIS
kernel, that is used to convolve one of the images in order to correct the 
effects of the variable PSF, was derived in regions around the brightest 
objects in the field. Since only a few field stars were present in the field of
view, we were forced to include the galactic nucleus in order to
properly evaluate the PSF and to obtain a good subtraction over the
whole field. Unfortunately, this method prevented us to discover the
most inner SNe. We have estimated the size of the non-detectability
region by simulations: bright point sources ($m_H=16.5$) were added to
the nuclear regions of the original images and then recovered after
subtracting with ISIS.  The size of the region where SNe cannot be
detected is given in Tab.~\ref{observ}.

Although the detection of very nuclear SNe (within the central $\sim
0.5\arcsec$, corresponding to about 200-600 pc for our galaxy sample)
is still impossible even with NICMOS, there is a significant
improvement with respect to ground-based observations, where strong
residuals prevent the detection of even bright SNe ($m_K<17$) within
the central $2\arcsec-3\arcsec$.  \\

\section{Expected number of SNe} \label{expectedsnsect}

According to the relation provided by Mannucci et al. (\cite{mannucci}), 
the expected number of SNe in each galaxy every 100 yr is given by:
\begin{equation} \label{firsn}
	SNe_{FIR}=(2.4 \pm 0.1) \ \frac{L_{FIR}}{10^{10}\ L_{\sun}} \ \frac{\mathrm{SN}}{100\ yr}
\end{equation}
as a function of the far-IR luminosity $L_{FIR}$ of the host galaxy.\\
For each galaxy we have computed the control time (the amount of time in which a SN is expected to 
be brighter than the detection limit in each of our images) 
by using the limiting magnitudes in Tab.~\ref{observ} and the mean H-band light curve for 
core-collapse SNe derived by Mattila \& Meikle (\cite{mattilameikle}).
The control time in each image is defined as the amount of time the H-band light curve 
of a SN, at the distance of the galaxy, is brighter than our detection limit.
Using the FIR flux of each galaxy and the derived control time, 
eq.~\ref{firsn} yields (after assuming no extinction):
\begin{displaymath}
	SN_{expt}(out)=26.8 \pm 4.9
\end{displaymath}
by using the limiting magnitude outside the subtraction residuals, and 
\begin{displaymath}
	SN_{expt}(in)=12.6 \pm 4.7
\end{displaymath}
after assuming the more conservative limiting magnitude in the nuclear
regions.  The uncertainty on the expected number of SNe is dominated
by the large intrinsic dispersion of the SN light
curves in the near-IR, while uncertainties due to the the limiting
magnitude estimates and the use of equation (\ref{firsn}) are
negligible. Therefore the error bars correspond to an upper and a lower 
limit on the number of events, as expected when
using as reference the light curves corresponding, respectively, to the brighter
and fainter envelope of the H-band SN light curves.
The control time and the number of expected events for the two images of each galaxy 
are listed in Tab.~\ref{expected}.
\begin{table}[ht!]	
	\begin{center}
	\begin{tabular}{l r c c}
	\hline \hline
	Galaxy & C-Time & Exp. SNe & Exp. SNe \\
	& (days) & $\textrm{Mag}_L(out)$  & $\textrm{Mag}_L(in)$ \\
	\hline
NGC34             &    155.6   &  0.49 &  0.22 \\ 
NGC 1614       	  &    132.6   & 0.70 & 0.22 \\	
VII-ZW031         &    79.4    & 1.56 & 0.45  \\
IRAS 05189-2524   &    10.8    & 1.56 & 0.05  \\
IRAS 08572+3915   &    156.4   & 1.25 & 1.00 \\
UGC 5101          &   98.6     & 1.60 & 0.51   \\
NGC 3256          &    331.8   & 0.95 & 0.60 \\
IRAS 10565+2448   &    121.8   & 1.38 & 0.59 \\
NGC 3690          &    311.8   & 1.61 & 1.08 \\
NGC 4418          & 428.6      & 0.38 & 0.28    \\
Mrk 273           &   120.2    & 2.40 & 0.99  \\
UGC 8782          &    153.2   & 4.04 & 1.88 \\
IRAS 14348-1447	  &    43.2    & 2.13 & 0.53  \\
Arp 220        	  &    254.8   & 3.68 & 2.21 \\
NGC 6090       	  &    158.6   & 0.52 & 0.23 \\
IRAS 20414-1651	  &    171.0     & 1.20 & 1.10   \\
IRAS 23128-5919	  &   103.2    & 1.35 & 0.43  \\
	\hline
	All galaxies: & 7.8 yr & 26.8 & 12.4 \\
	\hline \hline
	\end{tabular}
\caption{Expected SNe according to eq. \ref{firsn}, using the mean H-band light curve 
for core-collapse SNe derived by Mattila \& Meikle (\cite{mattilameikle}) as reference 
and assuming no extinction. We are reporting the expected number of events 
using both $\textrm{Mag}_L(out)$ and $\textrm{Mag}_L(in)$ limiting magnitudes (see text). \textit{C-Time} is the 
control time of our observations assuming the more conservative $\textrm{Mag}_L(in)$.}
\label{expected}
	\end{center}
\end{table}

\section{Results and Implications} \label{implications}

Despite the high number of SNe expected in the NICMOS imaging of our
sample, only a possible candidate was discovered in Arp 220 and it is discussed in Appendix. 
This corresponds to an observed rate over the full sample smaller than 
\begin{equation}
	\mathrm{SN}_{obs}<\frac{\mathrm{N}_{obs}^{max}(90\%)}{\mathrm{CT}(in)}=\frac{3.89\ \mathrm{SN}}{7.8\ \mathrm{yr}}=0.5\ \mathrm{SN/yr}
\end{equation}
where $\mathrm{N}_{obs}^{max}(90\%)$ is the maximum number of SNe compatible with one detection 
at 90\% confidence level, assuming Poisson statistics, and $\mathrm{CT}(in)$ is the total Control 
time for the more conservative $\textrm{Mag}_L(in)$ limiting magnitude. \\
The observed lack of SNe can be explained in several ways:
\begin{enumerate}
	\item The FIR flux is dominated by obscured Active Galactic
	Nuclei. In this case most of the FIR luminosity would not be
	related to the star formation, but it would be AGN heated. Indeed some of the
	sources in our sample do host an AGN, as inferred from X-rays
	and optical spectra (e.g. UGC 5101, Mrk 231, NGC
	3690). However, even in these cases most of the FIR luminosity
	appear to be dominated by the starburst component (Corbett et
	al. \cite{corbett}, Thean et al. \cite{thean}, Genzel et al.
	\cite{genzel}, Lutz et al. \cite{lutz}, Clements et
	al. \cite{clements}).

	\item Another possibility is based on the fact that underluminous 
	SNe (e.g. Pastorello et al \cite{pastorello}) may form a significant 
	fraction of all core-collapse events. If this is the case, these SNe would stay 
	above our detection limit for a shorter time, thus decreasing the total control 
	time in Tab.~\ref{expected}.  
	However, even assuming that all SNe in our galaxies are underluminous, 
	we still expect $\sim 22$ or $\sim 8$ SNe using $\textrm{Mag}_L(out)$ and 
	$\textrm{Mag}_L(in)$ limiting magnitudes respectively.		

	\item If most starburst activity is concentrated within the
	inner 200-300 parsec of the galaxy a large fraction of SN
	events would not have been detected in our images (last column
	of Tab.~\ref{observ}).  It was indeed shown (e.g. Petrosian \&
	Turatto 1990; Bressan et al. 2002) that active and star
	forming galaxies show a higher concentration of SNe toward the
	center than it is observed in normal galaxies.  As discussed
	in detail in Mannucci et al. (\cite{mannucci}), current data
	do not put tight constraints on the size of the starburst
	region in those galaxies.  Smith et al. (\cite{smith}), Rovilos et al. 
	(\cite{rovilos05}) and 
	Lonsdale et al. (\cite{lonsdale}) have monitored the central
	arcsec of the nucleus of Arp 220 with VLBI at 18 cm, with a
	resolution of a few milli-arcsec, and in principle they could
	detect the SNe inside our ``non-detection'' zone.  They revealed
	a few variable compact radio sources attributable to SN. They detected 
	a total of 9 new sources in 9 years of observations, thus providing 
	a radio SN rate of $\sim 1$ SN/yr. However, this value 
	is still uncertain, as core-collapse SNe show a broad range of radio luminosities 
	(e.g. Weiler et al. \cite{weiler05}). Similar results
	were obtained in NGC 3690 (Neff et al. \cite{neff}) and Mrk
	273 (Bondi et al.\cite{bondi}), finding a nuclear rate between
	0.5-1 SN/yr. After comparing these results with the
	``observed'' rate of $< 0.5$ SN/yr discussed in this paper, 
	or even with an observed rate of $<0.25$ SN/yr as derived using 
	$\textrm{Mag}_L(out)$ instead of $\textrm{Mag}_L(in)$ limiting magnitude,
	one can infer that at least some SNe have been lost in the IR 
	because they occurred too close to the nucleus. 
	However, the nuclear SN rate observed at radio wavelengths in these 
	galaxies is only the 25\% of the total rate expected from their 
	FIR luminosity (see Table~\ref{expected}).
	
	\item The most likely possibility is that many events are so
	embedded in dust to be highly extincted even in the
	near-IR. In order to compute the (average) $A_V$, we have dimmed the
	template light curve with different amount of extinction up to
	match one (or less) SN detection, at a formal confidence level
	of $90\%$.  We have obtained 
	$A_V>25$ and $A_V>11$, after
	using $\textrm{Mag}_L(out)$ and $\textrm{Mag}_L(in)$,
	respectively, and assuming the standard Galactic extinction law 
	(Rieke \& Lebofsky \cite{rieke85}). If we include the candidate SN in Arp
	220, the needed extinction decreases to $A_V>22$ and
	$A_V>9$. We noticed that such high extinctions are not
	unlikely to occur in these powerful sturbursting systems
	(e.g. Genzel et al. \cite{genzel}, Sturm et al. \cite{sturm}).

\end{enumerate}

\begin{figure*}
	\centering
	\resizebox{\hsize}{!}{\includegraphics{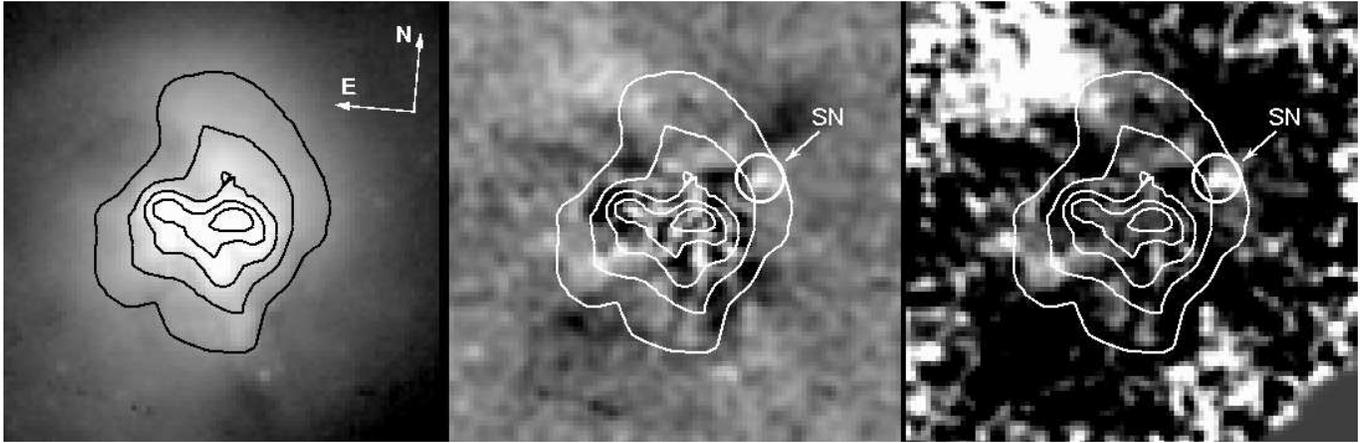}}
	\caption{\label{possn} \textit{Left Panel} - the discovery image of a candidate SN 
	in ARP 220, obtained  
	with NICMOS on 10th January 2004. The box is about $6" \times 6"$ \textit{Middle Panel} - 
	the electronic subtraction with the archive image obtain in 1997. The subtraction 
	was optimized by using a PSF matching algorithm. The distance between the SN and the 
	nucleus is only $\sim 1.1"$ (SN location: R.A. 15:34:57.13; DEC. +23:30:12.0; J2000). 
	\textit{Right Panel} - the subtraction divided by the 
	original image, to show the fraction of residual flux. The color scale spans between 
	$-1\%$ (black) to $3\%$ (white) of the galaxy flux. The candidate SN contains 
	about 4\% of the galaxy flux at that location.}
\end{figure*}

\section{Conclusions}	\label{conclusionsection}

We have presented the analysis of near-IR NICMOS-HST images of 17
galaxies, which were observed at two different epochs, with the goal
of detecting nuclear obscured SNe.  This study was prompted by the
shortage of SNe detected in near-IR monitoring campaigns of starburst
galaxies and by the possibility that most of the missing SNe occur in
the nuclear region, where the limited angular resolution of
ground-based observations prevents their detection.

Taking advantage from the high angular resolution of NICMOS and its
stable PSF, we were able to explore the central regions of
starbursting galaxies to search for obscured SNe, up to about
$0.5\arcsec$ from the nucleus and down to a limiting magnitude ranging
from $H\sim18.0$ in the very nuclear regions of the galaxies (closer
than $\sim1\arcsec$ to the nucleus) to $H\sim21$ further away from the
strongest subtraction residuals.  This is a clear improvement with
respect to ground based searches which are limited to $\sim2-3\arcsec$
from the nucleus. Within a radius of $\sim0.45\arcsec$ from the
center, even NICMOS is unable to reliably detect SNe. Our analysis
found a possible SN in Arp 220, but due to its faintness and proximity
to the nucleus we were not able to provide the spectroscopic
confirmation of this event. From our data we have derived a nuclear SN
rate $\leq 0.5$ SN/yr per galaxy. The comparison with rates measured
via radio-surveys suggests that some SNe occurring in the inner $\sim
500$ pc are lost. However, this is not enough to explain the
discrepancy between expected and observed number of SNe. We have
discussed various possible explanations, the most likely of which is
that in these starburst galaxies dust extinction is so high ($\rm
A_V\gtrsim11-25$) to be effective in obscuring SNe even in the
near-IR.

\appendix
 
\section{A possible Supernova in Arp 220}

A possible SN was discovered in ARP 220, the most active galaxy in our sample, 
observed on January 10th, 2004 (see Fig.~1).  It was found at $1.1\arcsec$ from the brightest
nucleus of the galaxy (SN location: R.A. 15:34:57.13; DEC. +23:30:12.0; J2000), and
it might be the most nuclear SN ever discovered.  It was detected as a
positive residual after careful alignment, normalization, reduction to
the same PSF and subtraction of two images of the galaxy taken at
different epochs. The SN was detected with a significance of 3.5
$\sigma$ with respect to the background noise, and it results in a
magnitude of $H\sim20.5$.
It is unlikely that the ``bright spot'' shown in Fig. \ref{possn} is the result of an
incorrect subtraction of the PSF. In fact, the luminosity of the SN 
candidate is about twice larger than the
contribution of the PSF of the bright galactic nucleus computed at the position
of the SN candidate. In addition, the candidate SN is
located away from the bright PSF spikes (that can not be perfectly
subtracted due to their highly asymmetric shape), where the limiting magnitude 
is much higher than the most conservative $\textrm{Mag}_L(in)$ reported in 
Tab.~\ref{observ}. In fact, our simulation show that a SN with $\textrm{H}=20.5$ would be 
recoverable close to the candidate SN location with comparable S/N.\\
The absolute
magnitude of the detected object is about M$_H=-13.8$, i.e. about 4
magnitudes fainter than an average type II SN at maximum light. This
may indicate that this SN has been discovered at about 300 days from
its maximum light, according to the template H-band light curve by
Mattila \& Meikle (\cite{mattilameikle}). Alternatively this SN could
have been discovered near its maximum light, but it is a very
extincted object, about 4 magnitudes in the H band, corresponding to
$A_V=23$.\\
  We have obtained spectroscopic follow-up of this object in the J band 
using ISAAC at the VLT, with the low resolution grism (R=500) and a 1$\arcsec$ slit, 
in order to detect the broad lines typical of
the nebular phase of a SN such as Pa$\beta$ and
[CaII]$\lambda$7300,8500 (see e.g. Maiolino et
al. \cite{maiolino}). The spectrum was obtained on April 11th, 2004, 3 
months after detection. 
Although the spectrum exposure time of 1h42' was chosen in order to have clear detection 
of the SN features, assuming a source up to $\mathrm{H}\sim21.5$, the signal to noise of the spectrum
was not sufficient to detect SN features over the bright galaxy spectrum.
The non-detection may still be due to 
the long delay time between the HST discovery image and spectroscopic observations. 
In fact, a SN is expected to became $\sim 2$ magnitudes fainter in the three months occurred between the 
two observations in the near-IR, corresponding to $H\sim22.5$. As a result the SN 
detection in ARP 220 remains tentative.

\begin{acknowledgements}
This research was supported in part by the National Science  Foundation under
Grant No. PHY99-0794
\end{acknowledgements}

\end{document}